\documentclass[a4paper,twocolumn,aps,prx,superscriptaddress,showpacs]{revtex4-1}
\usepackage{amsmath,amssymb,bm,graphicx,xcolor,hyperref}
\hypersetup{colorlinks,citecolor=blue,linkcolor=red,urlcolor=blue}

\begin{document}

\title{Phase transitions and correlations in fracture processes where
  disorder and stress compete}

\author{Santanu Sinha\email{santanu@csrc.ac.cn}}
\affiliation{Beijing Computational Science Research Center,
  10 East Xibeiwang Road, Haidian District, Beijing 100193, China.}
\affiliation{PoreLab, Department of Physics, Norwegian University of
  Science and Technology, NO--7491 Trondheim, Norway.}

\author{Subhadeep Roy\email{subhadeep.roy@ntnu.no}}
\affiliation{PoreLab, Department of Physics, Norwegian University of
  Science and Technology, NO--7491 Trondheim, Norway.}

\author{Alex Hansen\email{alex.hansen@ntnu.no}}
\affiliation{PoreLab, Department of Physics, Norwegian University of
  Science and Technology, NO--7491 Trondheim, Norway.}
\affiliation{Beijing Computational Science Research Center,
  10 East Xibeiwang Road, Haidian District, Beijing 100193, China.}

\date{\today}

\begin{abstract}
  We study the effect of the competition between disorder and stress
  enhancement in fracture processes using the local load sharing fiber
  bundle model, a model that hovers on the border between analytical
  tractability and numerical accessibility. We implement a disorder
  distribution with one adjustable parameter. The model undergoes a
  localization transition as a function of this parameter. We identify
  an order parameter for this transition and find that the system is
  in the localized phase over a finite range of values of the
  parameter bounded by a transition to the non-localized phase on both
  sides. The transition is first order at the lower transition and
  second order at the upper transition. The critical exponents
  characterizing the second order transition are close to those
  characterizing the percolation transition. We determine the
  spatiotemporal correlation function in the localized phase. It is
  characterized by two power laws as in invasion percolation. We find
  exponents that are consistent with the values found in that problem.
\end{abstract}

\pacs{62.20.mm, 64.60.av, 46.50.+a, 81.40.Np}

\maketitle

\section{Introduction}
\label{intro}

It has been known for a long time that heterogeneities make materials
more resilient against failure under load by offsetting the point at
which a microfracture becomes unstable \cite{wz17}. Examples of
materials where heterogeneities play an important role are concrete
\cite{mhk93, ldl19} and carbon-fiber composites \cite{ksl08} which
possess larger strength and toughness than the individual components
\cite{l92,m93}. Heterogeneities introduce spatial disorder in the
local material strength and create spatially dependent dynamic stress
field. Under sufficiently large external stress, a micro-crack appears
at the weakest point of the material and creates higher stress at the
crack tips. Whether this microcrack will grow to a catastrophic
failure or different micro-cracks will appear with increasing stress,
depends on the dynamic competition between local stress enhancement
and local material strength. Therefore, understanding the role of
disorder in detail is crucial for technological purposes, in order to
optimize the material strength.

\begin{figure*}[t]
  \centerline{\hfill\includegraphics[width=0.98\textwidth,clip]{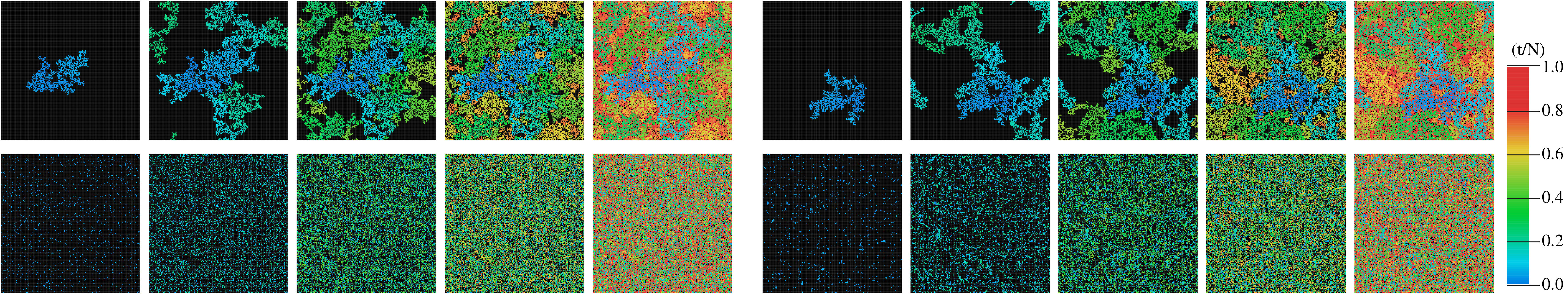}\hfill}
  \smallskip
  \centerline{\hfill $D>0$ \hfill \hfill $D<0$ \hfill}
  \caption{\label{figSnap} Growth of cracks in a bundle of
    $256\times 256$ fibers with threshold distribution from Eq.
    (\ref{eqPx}). The left and right panels correspond to $D>0$ and
    $D<0$ respectively. The top and bottom rows in each panel
    correspond to low ($|D|=0.02$) and high ($|D|=2.0$) disorders
    respectively. The black sites are intact fibers and the colored
    sites are broken fibers. The colors represent the breaking
    sequence ($\tau = t/N$) as indicated by the scale at right. The
    five snapshots in each set show the evolution of cracks at
    $\tau=0.04$, $0.2$, $0.4$, $0.6$ and $0.9$.}
\end{figure*}

The interplay between disorder and dynamical effects during the
breakdown process of brittle materials was the focus of much research
within the statistical physics community during the eighties and
nineties \cite{hr14}. This research was based on a number of
lattice-based models where the links would have a maximum load before
they would fail drawn from some spatially uncorrelated distribution.
For example the fuse model \cite{ahr85,dbl86,kbrah88} consisted of a
network of electrical resistors that would act as fuses, the central
force model consisted of a network of freely rotating Hookean springs
that would fail if the load they carried crossed a threshold
\cite{hrh89}, and the beam model \cite{hhr89} where the links in the
network would be elastic beams that would fail if certain criteria was
fulfilled \cite{sh19}.  We summarize the picture that emerged
qualitatively in the following.  There are two reasons for a failure
to appear locally in a disordered material. Either it is due to the
material being weak at that local point or it is due to the stress
being high there.  Imagine now loading the disordered material.  At
the beginning of the breakdown process, the material will fail where
it is weak as there are no --- or few --- points with high
stress. When local failures develop, spots with intense stress appear
at the tips of microcracks.  As the failure process proceeds, these
high-stress spots will start dominating.  As it was argued by Roux and
Hansen, \cite{rh90} disorder makes local failures {\it repulsive\/}
whereas the stress field makes them {\it attractive.\/} Imagine that
there is a first local failure.  Draw a sphere around this first local
failure and identify the weakest spot within the sphere.  The larger
the sphere is, the weaker the weakest spot within it will be.  As a
result, if the next local failure is due to the local weakness of the
material, it will occur as far as possible from the first one.  Hence,
the failures are repulsive when they are caused by the disorder in the
strength of the material.  The effect of the stress field is
opposite. The closer one is to a local failure, the higher the largest
stress will be.  This makes it more likely that the next failure will
be close to the one that just appeared.  Hence, local failures attract
each other when they are due to high stress field. The result of this
is that there is a competition between disorder and stress
concentration throughout the failure process.  Early in the failure
process the disorder tends to dominate, resulting in local failures
appearing distributed throughout the material.  However, as the stress
field takes over, there is localization ending with a single growing
crack beginning to dominate.  Depending on the disorder, localization
may occur early or later in the process: The wider the disorder, the
later on the onset of localization would occur.  In the limit of
infinite disorder \cite{rhhg88,moh12,szs13}, localization never sets
in and the failure process is a {\it screened\/} percolation --- i.e.\
a process where links fail at random given that they are connected in
such a way that they carry stress.  When the disorder is weak enough
to cause a competition with the stresses, a phase diagram may be
constructed showing the onset of localization as a function of the
disorder \cite{ah89,ahhr89,hhr91}.

\begin{figure}[b]
  \centerline{\hfill\includegraphics[width=0.48\textwidth,clip]{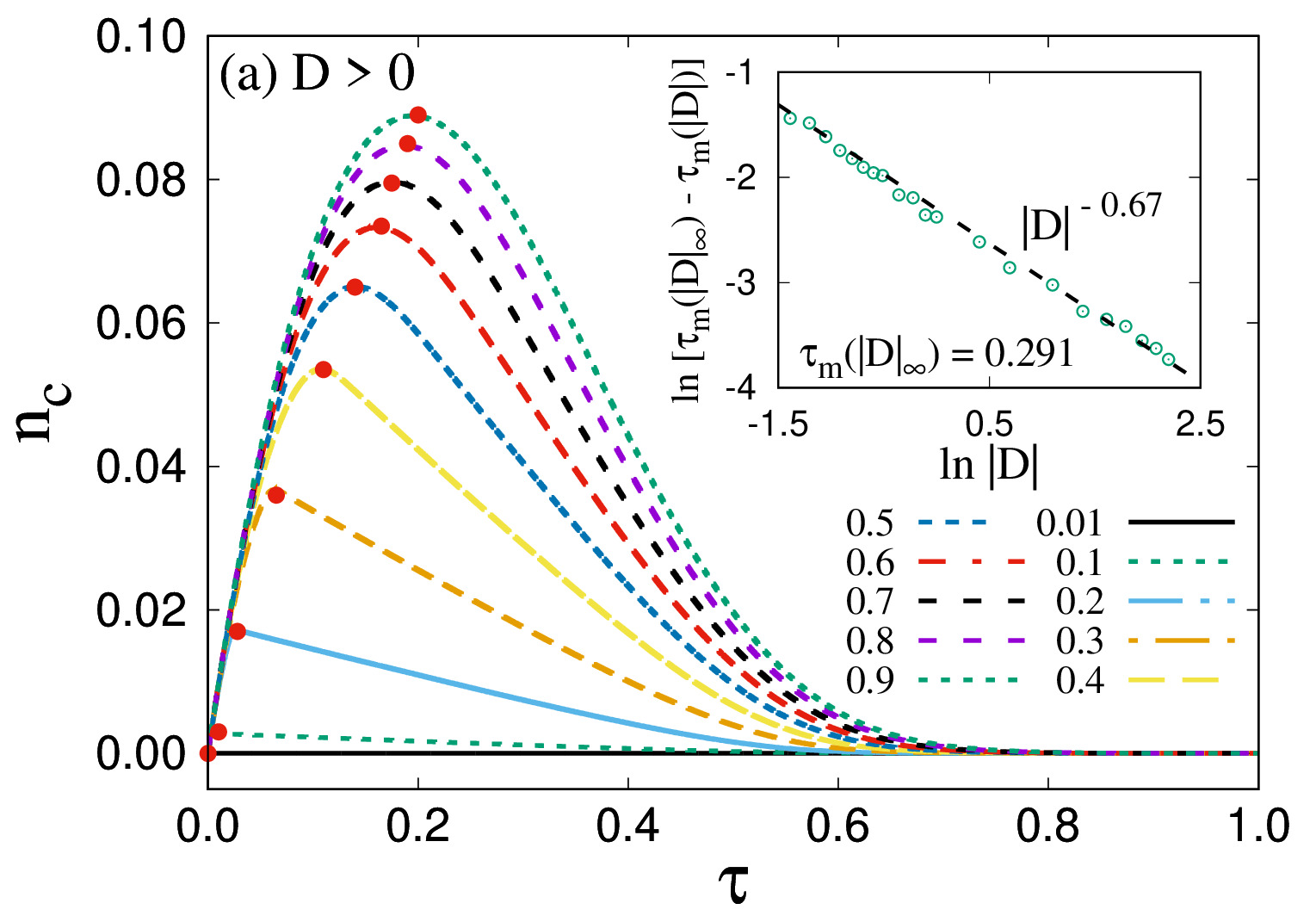}\hfill}
  \centerline{\hfill\includegraphics[width=0.48\textwidth,clip]{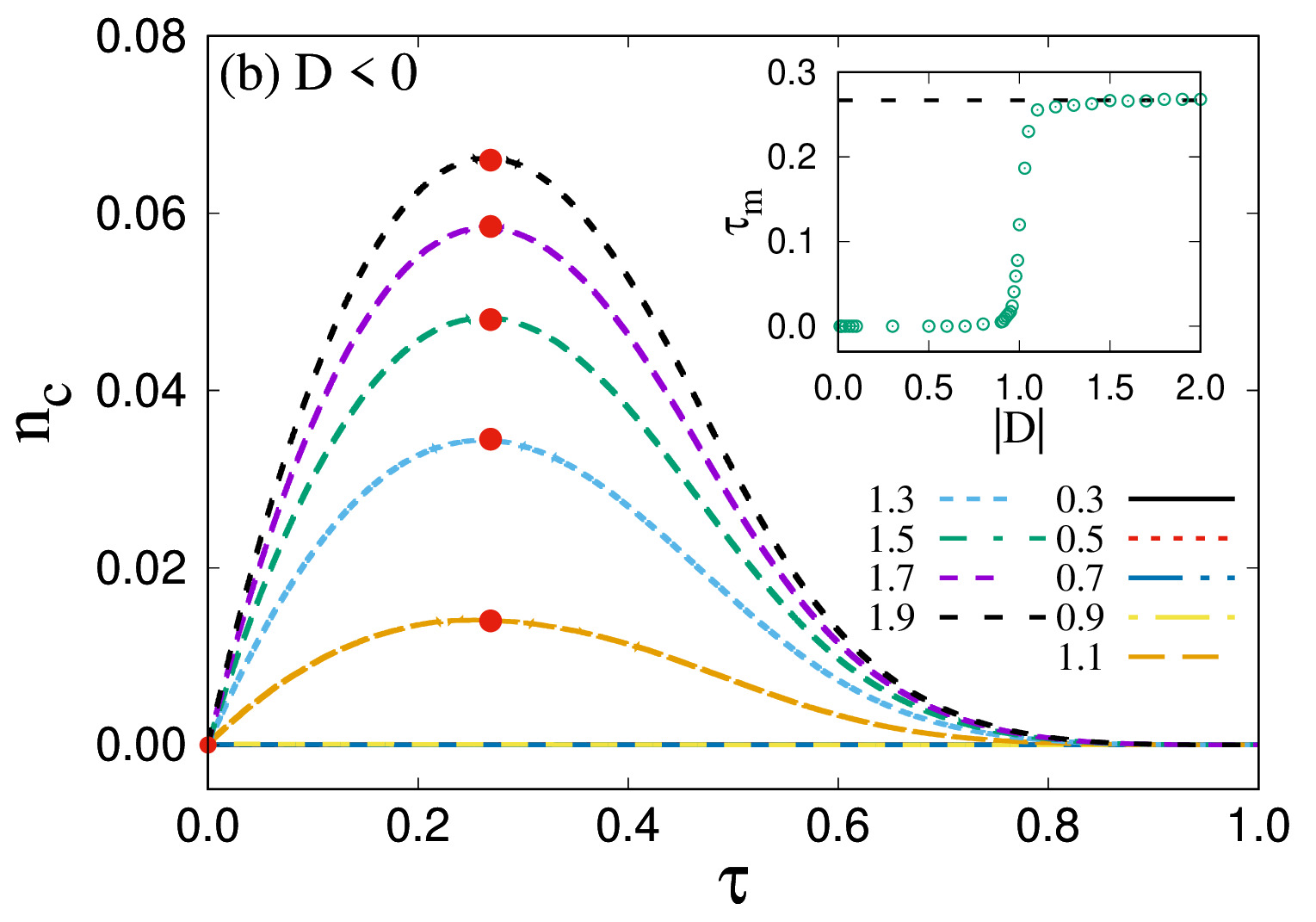}\hfill} 
  \caption{\label{fignc} Variation of crack densities $n_c$ with
    fraction of broken bonds $\tau$ for $L=256$ at different
    disorders. The $|D|$-values are indicated in the legends. The red
    dots indicate the peaks at ($\tau_{\rm m},n_c^{\rm max}$). For
    $D>0$ we see a gradual increase in $\tau_{\rm m}$ with increasing
    $|D|$ whereas for $D<0$ we first see a discontinuity from zero to
    a finite value and then it remains almost constant at
    $\approx 0.27$. The inset of (a) shows the scaling
    $\tau_m^+(|D|_\infty)-\tau_m^+(|D|) \sim |D|^{-\gamma}$ for $D>0$
    where $\gamma = 0.67\pm 0.01$ and
    $\tau_m^+(|D|_\infty) = 0.291\pm 0.011$. The inset of (b) shows
    the convergence of $\tau_m^-$ towards $|D|\to\infty$ for $D<0$.}
\end{figure}

It is the aim of this work to study the effect of the competition
between disorder and stress enhancement using the {\it fiber bundle
  model\/} \cite{phc10,hhp15}.  The advantage of using this model over
other models is that it is not computationally demanding, leading to
good statistics for large samples.  Furthermore, it has a high level
of analytical tractability.  We describe the fiber bundle model in
detail in Sec. \ref{FBM}. It should be noted that Stormo et al.\
\cite{sgh12} studied the interplay between disorder and stress
enhancement in the {\it soft clamp model\/} which is a more complex
version of the fiber bundle model than the version we study
here. Their conclusions differ from those we present here; this due to
a very different way of analyzing the fracture process.  We return to
this work in the concluding section. We implement a disorder
distribution that has one adjustable parameter.  By tuning this
parameter, we produce disorders with power law tails towards either
zero strength or infinite strength.  We focus in Sec. \ref{loc} on the
localization transition that occurs for some value of the disorder
parameter, see Fig.  \ref{fignc}.  For the range of the values of the
disorder parameter that produces power law tails towards zero
strength, we find a second order phase transition, whereas for the
range of values for the disorder parameter that produces a power law
tail towards infinite strength, we find a first order phase
transition.  We determine the values of critical exponents associated
with the second order phase transition. They are close to those found
in percolation. In Sec.  \ref{corr} we study the spatiotemporal
correlation function first introduced by Furuberg et al.\ \cite{ffa88}
in connection with invasion percolation.  We determine the scaling
exponents characterizing the correlation function and find them to be
close to the values observed for invasion percolation.  We relate the
exponents to other exponents characterizing the transition using
theory developed by Roux and Guyon \cite{rg89}, Maslov \cite{m95} and
Gouyet \cite{g90}. We end by drawing our conclusions in
Sec. \ref{conc}.

\section{Description of the Model}
\label{FBM}

The fiber bundle model is a model of fracture where one can control
the range of stress enhancement upon the appearance of a crack.  In
this model, $N$ fibers (Hookean springs) are placed between two clamps
under external force $F$. Each fiber carries a force
\begin{equation}
  \displaystyle
  f = \kappa x\;,
  \label{eqfkx}
\end{equation}
where $\kappa$ and $x$ are respectively the elastic constant and
extension of the fiber. The extension of each fiber has a threshold
($x_t$), beyond which it fails and the load it was carrying is
distributed among surviving fibers according to some pre-set rule.
The stress distribution scheme models the local stress enhancement
whereas the disorder in $x_t$ models the local material
heterogeneity. If the load of the failed fibers is distributed over
all the surviving fibers, there is no local stress enhancement. This
is the {\it equal load sharing\/} (ELS) scheme.  The ELS fiber bundle
model was introduced by Peirce in 1926 \cite{p26} as a simple model
for failure in fibrous materials. Daniels approached the ELS fiber
bundle model as a problem in statistics in a seminal paper in 1945
\cite{d45}. If the load is distributed only to the nearby fibers, we
are dealing with the {\it local load sharing} (LLS) model \cite{hp78,
  hp91}.  Here local stress enhancement competes with the local
heterogeneity leading to localization.  Sornette introduced the ELS
fiber bundle model to the statistical physics community in 1992
\cite{s92}.  Soon, the focus of this community was on the rich
avalanche statistics that this model offer
\cite{hh92,hh94,zd94,khh97}, which in the ELS case is analytically
tractable.

The sequential --- or time --- correlation between failure events in
the fiber bundle model makes it possible to explore the brittle to
ductile transition \cite{asl97} within it, a well-studied phenomenon
in material science. The spatial correlations between the failures, on
the other hand, have not been studied in detail.

In the LLS fiber bundle model, the load carried by the failed fibers
is distributed equally among their nearest intact neighbors.  We
define a {\it crack\/} as a {\it cluster\/} of $s$ failed
nearest-neighbor fibers. The {\it perimeter\/} of the crack is the set
of $p$ intact fibers that are nearest neighbors to the failed fibers
constituting the crack. These nearest neighbors define the {\it
  hull\/} of the cluster \cite{ga87}.  The force on an intact fiber
$i$ at any instance is then give by
\begin{equation}
  \displaystyle
	f_i = \sigma\left(1 + \sum_{J(i)}\frac{s_{J(i)}}{p_{J(i)}}\right)\;,
  \label{eqlls}
\end{equation}
where $\sigma=F/N$, the force per fiber. The summation runs over all
cracks $J(i)$ that are neighbors to fiber $i$.  This redistribution
scheme is independent of the lattice topology and it is also
independent of the failure history \cite{skh15}, that is, the complete
stress field can be calculated from the present arrangement of intact
and broken fibers without having to take into account the order in
which the fibers failed.

Suppose $t$ fibers have failed. We determine which fiber will fail at
time $t+1$ in the following way \cite{hhp15}. Let $f_i^1$ be the force
on fiber $i$ if we set the average force on the fibers $\sigma=1$. We
then calculate
\begin{equation}
  \label{criterion}
  \lambda(t+1)=\max_{i}\left(\frac{f_i^1}{\kappa x_{t,i}}\right)\;,
\end{equation}
which denotes the fiber $j(t+1)$ that fails at time $t+1$. The force
$\sigma=\sigma(t+1)$ at which this force fails is given by
\begin{equation}
\label{sigmafrac}
	\sigma(t+1)=\frac{1}{\lambda(t+1)}\;.
\end{equation}

The failure thresholds $x_{t}$ of the fibers are assigned by
generating a random number over unit interval and raising it to power
$D$ which corresponds to the cumulative distribution \cite{hhr91,sh19}
\begin{equation}
  \displaystyle
  P(x_{\rm t}) = \begin{cases}
	  x_{\rm t}^{1/|D|}\;, x_t\in[0,1] & \text{when} \;  D>0\;, \\
	  1-x_{\rm t}^{-1/|D|}\;, x_t\in[1,\infty)  & \text{when} \;  D<0\;.
  \end{cases}
  \label{eqPx}
\end{equation}
This threshold distribution allows us to control the disorder by the
value of $D$, higher value of $|D|$ implies higher
disorder. Furthermore, $D>0$ and $D<0$ respectively correspond to the
distributions with power law tails towards weaker and stronger fibers,
which, as we will see, make the failure dynamics very different.

\begin{figure}[t]
 \centerline{\hfill\includegraphics[width=0.48\textwidth,clip]{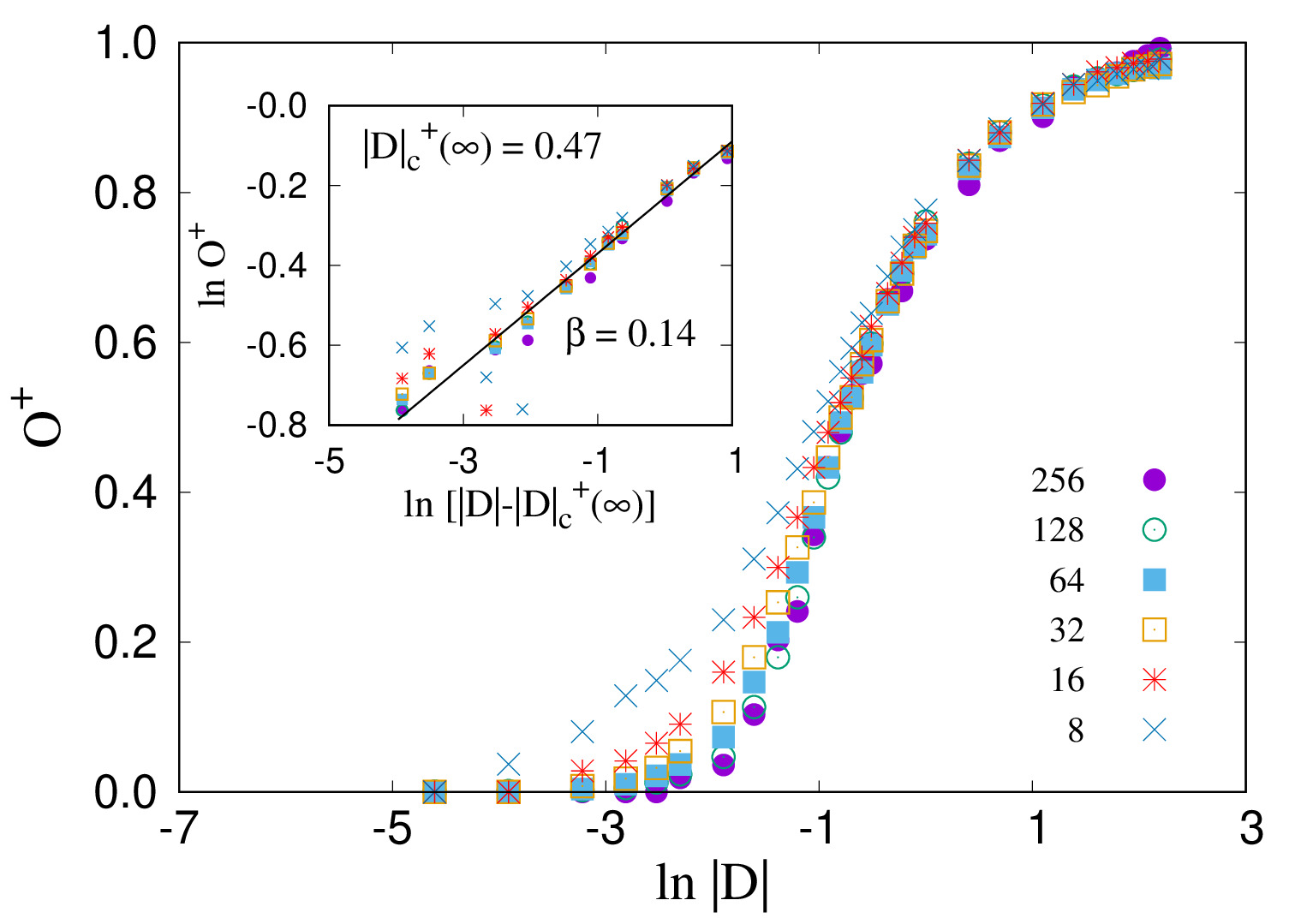}\hfill}
 \caption{\label{figOp}Plot of order parameter $O^+$ as a function of
   disorder $|D|$ for $D>0$. Different symbols indicate different
   system sizes $L$ where data are averaged over $200$ to $100000$
   samples depending on $L$. In the inset, we plot the scaling
   relation given in Eq. (\ref{eqOpSc}) where we find
   $|D|_{\rm c}^+(\infty) = 0.47\pm 0.04$ and $\beta = 0.14\pm 0.05$.}
\end{figure}

\section{Localization Transition}
\label{loc}
We implement the model on a square lattice of size $N=L\times L$
fibers. The growth of cracks for $L=256$ at different disorders are
shown in Fig. \ref{figSnap} where we see two distinct regimes, a
single crack growth at low disorder (top row) and random failures at
high disorder (bottom row). The color scale in this figure represents
where in the breaking sequence a given fiber fails, parameterized by
$\tau=t/N$, which we will discuss in detail later. At very low
disorder, the stress enhancement at the crack perimeter wins over the
fiber strengths and the crack always grows from the perimeter of the
existing crack. The local disorder at the perimeter makes the crack
grow as an invasion percolation cluster \cite{ww83}.

At high disorder, the strength of fibers win over the stress
enhancement and we see random clusters of failed fibers appear.  These
two cases appear to be same for both $D>0$ and $D<0$, however, a
completely different picture emerges between these two regimes when we
observe crack growth at intermediate disorders. In Fig. \ref{fignc},
we plot the crack density $n_c=N_c/N$, where $N_c$ is the number of
cracks (clusters) when $t$ fibers have broken, for different disorders
as a function of $\tau=t/N$. We see that $n_c$ stays close to $0$
($N_c=1$) for the whole failure process for small $|D|$ values whereas
for large $|D|$, $n_c$ increases with $\tau$ and reaches a maximum
($n_c^{\rm max}$), beyond which, the cracks start to coalesce as more
fibers are broken.  The position of the peaks at $\tau$-axis, marked
by the red dots in Fig. \ref{fignc}, vary continuously with disorder
for $D>0$ whereas the peak stays at almost same position for $D<0$
after a discontinuity between the low and high $|D|$ values. Moreover,
for two disorders with same values of $n_c^{\rm max}$, the peak
appears much earlier for $D>0$ compared to $D<0$. This shows that
individual cracks appear randomly for $D>0$ whereas cracks grow in
size together with the appearance of new cracks for
$D<0$. Intuitively, when the threshold distribution has a power-law
tail towards strong bonds ($D<0$), probability to find weak bonds at
existing crack perimeters are higher, which makes existing cracks to
grow. Whereas when power-law tail is towards the weak bonds ($D>0$),
probability to find strong bonds at the perimeter is high and new
cracks appear at different positions than the perimeter.

\begin{figure}[t]
  \centerline{\hfill\includegraphics[width=0.48\textwidth,clip]{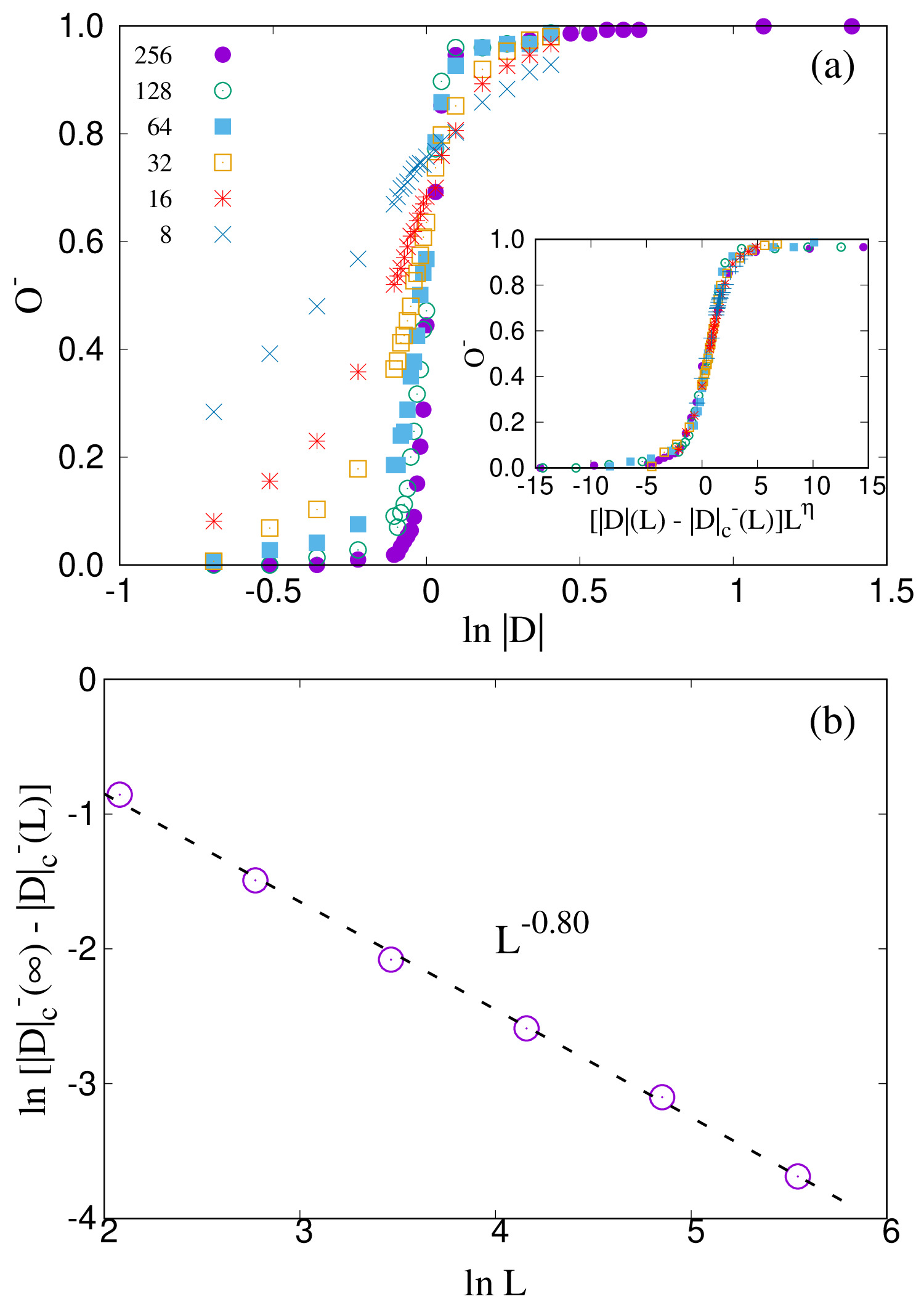}\hfill}
  \caption{\label{figOn} Plot of order parameter $O^-$ as a function
    of disorder for $D<0$ is plotted in (a) which shows abrupt
    transition with the increase in system size $L$. The inset in (a)
    shows the scaling with different $L$ according to Eq.
    (\ref{eqOnSc}). In (b), we plot
    $|D|_{\rm c}^-(\infty) - |D|_{\rm c}^-(L)\sim L^{-\theta}$ where
    we find $|D|_{\rm c}^-(\infty) = 1.03 \pm 0.02$ with
    $\theta = 0.80 \pm 0.01$.}
\end{figure}

To characterize this {\it localization transition,\/} we define an
{\it order parameter},
\begin{equation}
  \displaystyle
  O^{-/+}(|D|,L) 
	= \frac{\tau_{\rm m}^{-/+}(|D|,L)}{\tau_{\rm m}^{-/+}(|D|_\infty,L)}
    \label{eqOrd}
\end{equation}
where the plus and minus signs correspond to $D>0$ and $D<0$
respectively. Here $\tau_{\rm m}^{-/+}(|D|,L)$ is the fraction of
broken fibers at $n_c=n_c^{\rm max}$ and
$\tau_{\rm m}^{-/+}(|D|_\infty,L)$ is the value of $\tau_{\rm m}$ for
$|D|\to \infty$, i.e., when the breakdown is an uncorrelated
percolation process.  The measurements of
$\tau_{\rm m}^{-/+}(|D|_\infty,L)$ for $L=256$ are shown in the insets
of Fig. \ref{fignc}. For $D>0$ we observe the scaling,
\begin{equation}
\tau^+_{\rm m}(|D|_\infty)-\tau^+(|D|) \sim |D|^{-\gamma}\;,
	\label{scaling_tau_m}
\end{equation}
where we have that $\gamma=0.67 \pm 0.01$ and
$\tau^+(|D|_\infty)=0.291 \pm 0.01$. For $D<0$, $\tau_m^-(|D|)$
converges to a maximum value more abruptly from which we find
$\tau^-(|D|_\infty)=0.27$ as shown by the dashed line in the inset.

\begin{figure*}[t]
  \centerline{\hfill
    \includegraphics[width=0.32\textwidth,clip]{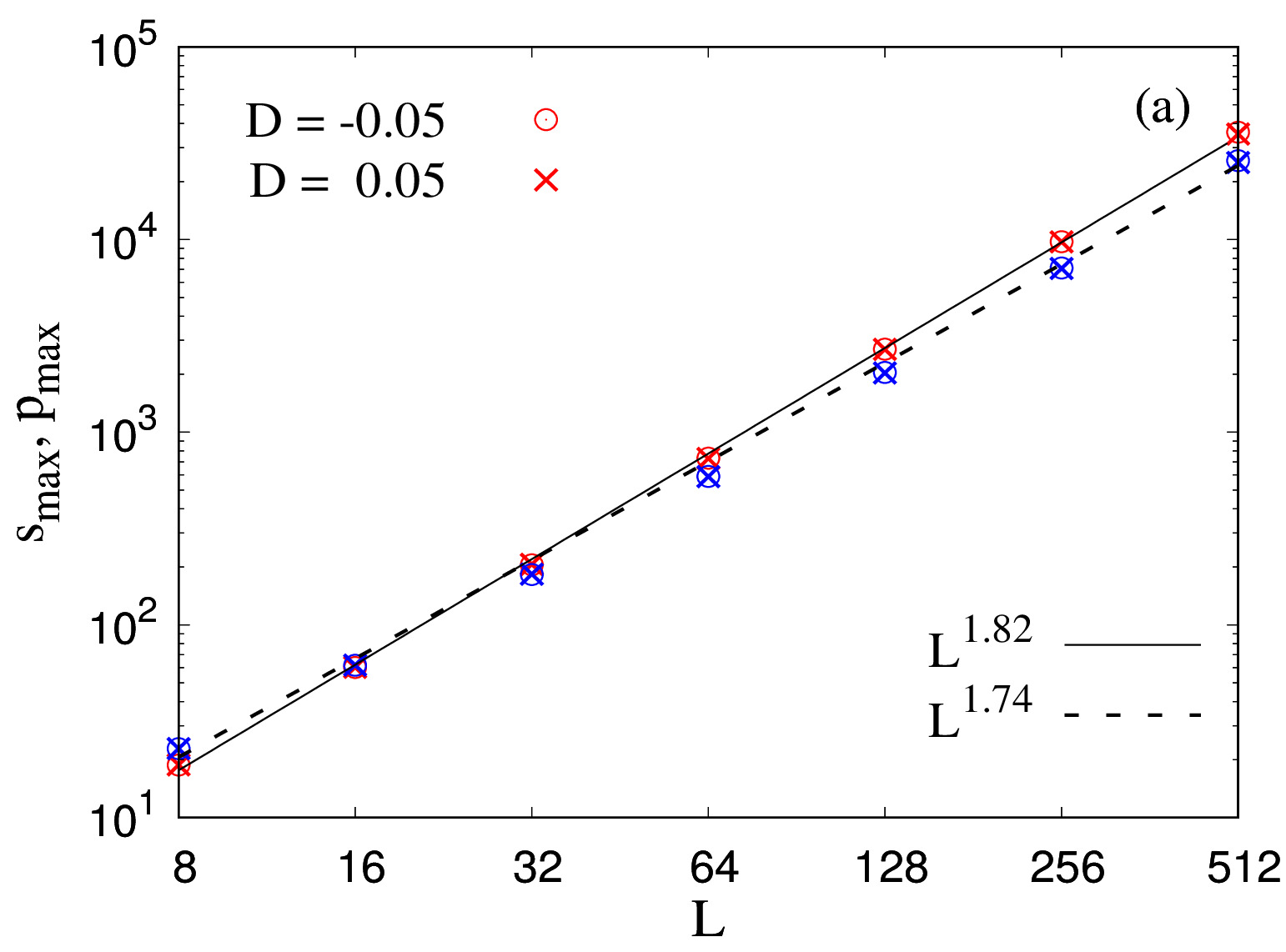}\hfill
    \includegraphics[width=0.32\textwidth,clip]{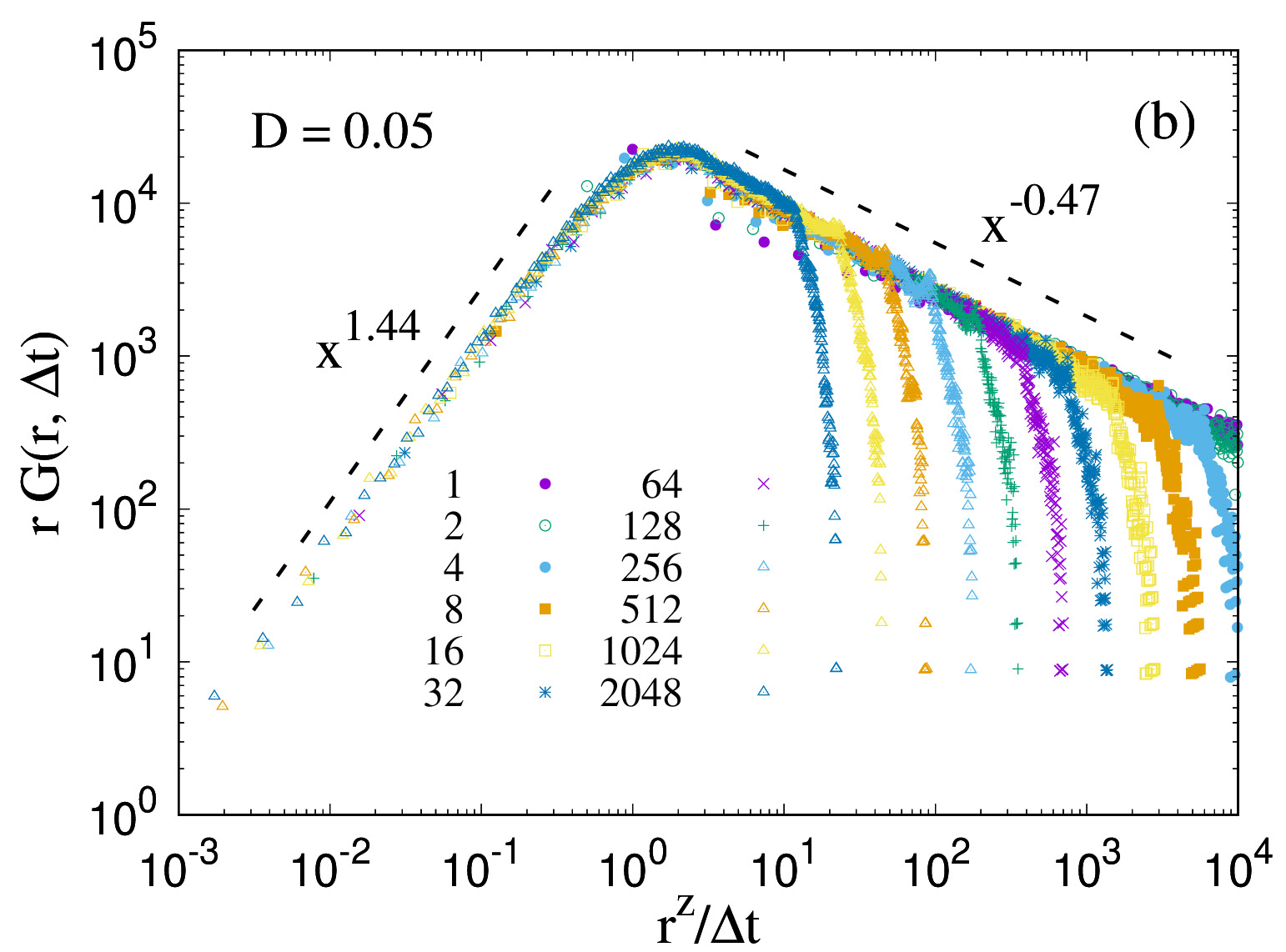}\hfill
    \includegraphics[width=0.32\textwidth,clip]{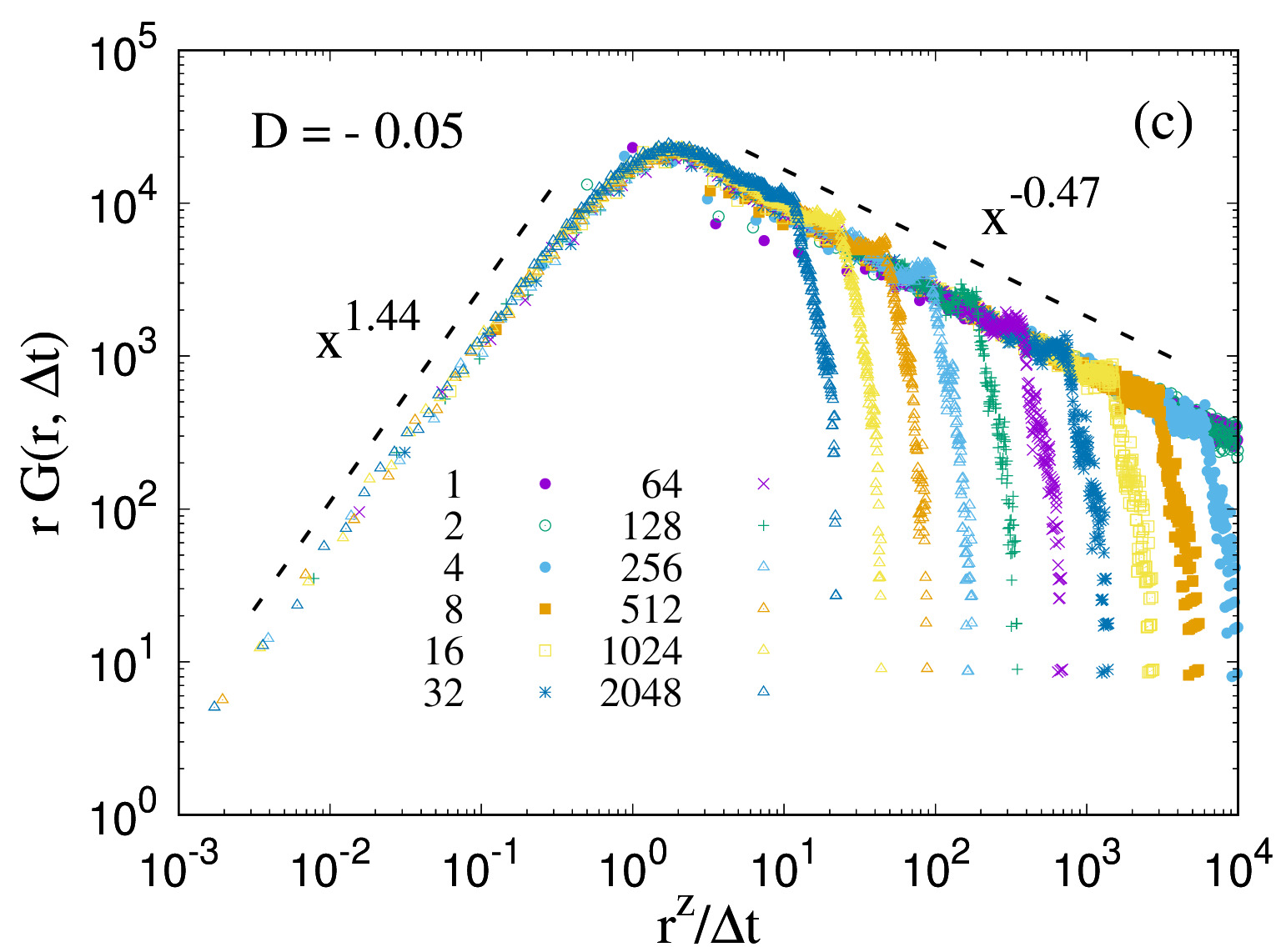}\hfill}
  \caption{\label{figCFn} (a) Plot of the largest crack size
    $s_{\rm max}$ (red symbols) and the largest perimeter size
    $p_{\rm max}$ (blue symbols) as a function of the system size
    $L$. From the slopes we find $d_{\rm f}=1.82\pm 0.02$ and
    $d_{\rm h}=1.74\pm 0.02$. In (b) and (c), we plot the pair
    correlation function (Eq.  (\ref{eqCFn})) for the $D = +0.05$ and
    $-0.05$ respectively, where we used $z=d_{\rm f}$ obtained from
    (a). The different plots correspond to different values of
    $\Delta t$ as indicated in the figures. From the two power law
    scalings we find the exponents $a = 1.44\pm 0.04$ and
    $b = 0.47\pm 0.06$.}
\end{figure*}

The transition for $D>0$ is explored in Fig. \ref{figOp}, where we
plot $O^+(|D|,L)$ as a function of disorder for different system
sizes. Here $O^+$ vary continuously from $1$ to $0$ similar to a
second order phase transition. We find the following scaling for
$O^+$,
\begin{equation}
  \displaystyle
  O^+(|D|,L) \sim \left[|D|(L) - |D|_{\rm c}^+(\infty)\right]^\beta
    \label{eqOpSc}
\end{equation}
where $|D|_{\rm c}^+(\infty) = 0.47 \pm 0.04$ and
$\beta \approx 0.14 \pm 0.05$. The exponent $\beta$ is close to
$5/36$, the order parameter exponent for percolation in two dimension
\cite{n87,c87}. However, $1/8$, the exact value from Onsager solution
of the two-dimensional Ising model \cite{o44}, is also within the
precision with which we know $\beta$.

The transition for $D<0$ is presented in Fig. \ref{figOn} where we
plot $O^-(|D|,L)$ as a function of disorder. Similar to a first order
phase transition, $O^-$ shows sharp transition from $0$ to $1$ as the
system size $L$ is increased. The plots for different $L$ can be
scaled as \cite{clb86},
\begin{equation}
  \displaystyle
  O^-(|D|,L) \sim \Phi\Big\{ \left[|D|(L)-|D|_{\rm c}^-(L)\right]L^\eta\Big\}
  \label{eqOnSc}
\end{equation}
with $\eta = 0.7$. This is shown in the inset of Fig. \ref{figOn}(a).
The transition disorder $|D|_{\rm c}^-(L)$ is observed to be a
decreasing function of $L$. To find the value of $|D|_{\rm c}^-$ as
$L\to \infty$, we use the scaling
$|D|_{\rm c}^-(\infty) - |D|_{\rm c}^-(L)\sim L^{-\theta}$ with
$|D|_{\rm c}^-(\infty) = 1.03 \pm 0.02$ and $\theta = 0.80 \pm
0.01$. This is shown in Fig. \ref{figOn}(b).

\section{Spatiotemporal Correlations}
\label{corr} 
We now turn to the investigation of the spatial and temporal
correlations during the breakdown process. First we measure the
fractal dimension of single cracks at low disorder, defined as
$s_{\rm max}\sim L^{d_{\rm f}}$ where $s_{\rm max}$ is size of the
largest incipient infinite cluster. This is shown in Fig. \ref{figCFn}
(a). We find $d_{\rm f}=1.82\pm 0.02$, and note that this close to the
fractal dimension of invasion percolation cluster with trapping
\cite{ww83, lz85}.  The colors in Fig. \ref{figSnap} represent the
sequence of fiber breaking and together with the positions of broken
fibers, they show the spatiotemporal map of the failure process. For
random failures at high disorder (bottom row), the pixels (each
representing a fiber) of different colors are randomly mixed, whereas
for the single crack growth (top row), the different colors are
clustered, indicating localization and hence spatial and temporal
correlations. To quantify this correlation, we measure {\it pair
  correlation function\/} $G(r,\Delta t)$, defined as follows: if a
fiber at position $\bm{r_0}$ breaks at a time $t$, then
$G(r,\Delta t)$ provides the probability that another fiber at
$\bm{r_1}$ will break at time $t+\Delta t$, where
$r=|\bm{r_1}-\bm{r_0}|$. Here time $t$ is measured in terms of the
number of broken fibers. This correlation function was first
introduced by Furuberg {\it et al.\/} \cite{ffa88} for invasion
percolation.  We assume in the following that the $G(r,\Delta t)$ has
the scaling form \cite{ffa88}
\begin{equation}
  \displaystyle
  G(r,\Delta t) = r^{-1}g\left(\frac{r^z}{t}\right)\;,
  \label{eqCFn}
\end{equation}
where $z$ is the dynamic exponent. In Fig. \ref{figCFn}, we plot
$rG(r,\Delta t)$ for the single crack growth regime as a function of
$r^z/t$ for $D>0$ (b) and $D<0$ (c). For $z=d_{\rm f}$, the
single-parameter function $g$ shows power-law behavior in both the
large- and small-argument limits,
\begin{equation}
  \displaystyle
  g(u) \sim
  \begin{cases}
    u^a  , & u \ll 1 \; \; \textnormal{(long time range)}\;,\\
    u^{-b}, & u \gg 1 \; \; \textnormal{(short time range)\;,}
  \end{cases}
  \label{eqPow}
\end{equation}
with a peak at $u=1$ or at $t=r^{d_{\rm f}}$ as shown in the
figure. This implies that the most probable growth of the crack after
time $t$ occurs at a distance $r^{d_{\rm f}}$, or, the most probable
growth at a distance $r$ occurs after a time $t^{1/d_{\rm f}}$. The
exponent $a$ in the long time range is found to be $1.44 \pm 0.04$ as
shown in Fig. \ref{figCFn}. This matches with the exponent for
invasion percolation dynamics \cite{ffa88}. Recently, this scaling was
observed experimentally for two-phase flow in porous medium during
slow drainage \cite{mmf17} with the exponent $a$ ranging between
$1.44$ and $1.73$. For the power law in the small-argument limit, we
obtain the exponent $b = 0.47 \pm 0.06$, which is different from
$0.68$ obtained by Furuberg {\it et al.\/} for invasion percolation
dynamics \cite{ffa88}.

The exponents and the errors mentioned above are obtained from the
least square fitting of the numerical data. The plots with multiple
data file, such as figure \ref{figOp} inset, figure \ref{figCFn} (b)
and (c), the error mentioned is the average of the errors
corresponding to individual data.

We may relate the exponents $a$ and $b$ to the exponents controlling
the avalanche structure of the failure process following Roux and
Guyon \cite{rg89}, Maslov \cite{m95} and Gouyet \cite{g90}.  The
minimum force applied to the fiber bundle for fiber number $t+1$ to
fail is $\sigma(t+1)$ given in Eq. (\ref{sigmafrac}).  We define a
{\it forward\/} avalanche (or burst) of size $\Theta$ starting at time
$t$ to be \cite{hh92,m95}
\begin{equation}
	\label{forwardburst}
	\sigma(t+k)<\sigma(t)\ {\rm for}\ k<\Theta\ {\rm and}\ 
	\sigma(t+\Theta)\ge\sigma(t)\;.
\end{equation}
A {\it backwards\/} avalanche (or burst) we define as
\begin{equation}
	\label{backwardsburst}
	\sigma(t-k)<\sigma(t)\ {\rm for}\ k<\Theta\ {\rm and}\
	\sigma(t-\Theta)\ge\sigma(t)\;.
\end{equation}

Roux and Guyon \cite{rg89} define two distributions, $P_{\Delta t}$
and $Q_\Theta(r)$. The first one, $P_{\Delta t}$, gives the
distribution of the smallest avalanches $\Theta$ that pass through
$\sigma(t)$ and $\sigma(t-\Delta t)$ averaged over $\sigma(t)$. It
obeys the power law,
\begin{equation}
\label{pdeltat}
	P_{\Delta t}(\Theta) \propto \frac{1}{\Delta t}
	\left(\frac{\Theta}{\Delta t}\right)^{-c_b}
	H(\Theta-\Delta t)\;,
\end{equation}
where $H(x)$ is the Heaviside function which is one when $x>0$ and
zero otherwise. The second one, $Q_\Theta$, gives the distribution of
distances between failing fibers within an avalanche of size
$\Theta$. Roux and Guyon assume it to follow the power law
\begin{equation}
	\label{qthetar}
	Q_\Theta(r)\propto\frac{1}{\Theta^{1/d_{\rm f}}}
	\left(\frac{r}{\Theta^{1/d_{\rm f}}}\right)^{\zeta}\;.
\end{equation}
The spatiotemporal correlation function $G(r,\Delta t)$ may then be
constructed from these two probability distributions,
\begin{equation}
\label{eq2}
G(r,\Delta t) = \int_{0}^{\infty} P_{\Delta t}(\Theta) 
	Q_{\Theta}(r) d\Theta\;,
\end{equation}
which integrates to
\begin{equation}
	\label{gcases}
	r G(r,\Delta t)\propto \begin{cases}
		\left(\frac{r^{d_{\text f}}}{\Delta t}\right)^{(1+\zeta)/d_{\text f}}
		& \quad \text{if} \ \ r^{d_{\text f}}<\Delta t\;,\\
		\left(\frac{r^{d_{\text f}}}{\Delta t}\right)^{1-c_b}
		& \quad \text{if} \ \ r^{d_{\text f}}>\Delta t\;.\\
		               \end{cases}
\end{equation}
Hence, we have in Eq. (\ref{eqPow}) that $a=(1+\zeta)/d_{\text f}$ and
$b=c_b-1$.  Roux and Guyon suggest $\zeta=d_{\text f}-1$ making $a=1$.
We propose here that $\zeta=d_{\rm h}$, the fractal dimension of the
hull \cite{f88}, defined as the set of the sites that are connected to
the cluster and at the neighbor to the surroundings. In the LLS fiber
bundle model, the stress of a crack is re-distributed to its
perimeter, the neighboring unbroken fibers of the crack. In Fig.
\ref{figCFn}, we plot this largest perimeter size $p_{\rm max}$ with
the system size $L$ and from the slope, we find
$d_{\rm h}=1.74 \pm 0.02$. This value matches with the value of
$d_{\rm h}$ for percolation hull, known to be $7/4$ by the relation
with $\nu$ by $d_{\rm h}=1+1/\nu$, where $\nu$ is the correlation
length exponent which is equal to $4/3$ in two-dimensional percolation
\cite{srg85, rgs86}.  Hence, we find that
\begin{equation}
	\label{aconjecture}
	a=\frac{1+d_{\rm h}}{d_{\text f}}\;.
\end{equation}
With $d_{\text f}=1.82\pm 0.02$ and $d_{\rm h}=1.74\pm 0.02$, we find
$a=1.51\pm 0.04$, which is close to the observed value.

Maslov \cite{m95} related the backwards avalanche exponent $c_b$ to
the exponent $c_f$ that governs the probability to find a forward
avalanche of size $\Theta$ when the stress is $\sigma_0$ via the
expression
\begin{equation}
\label{maslov}
	c_b=3-c_f\;.
\end{equation}
Furthermore, Goyuet in turn related $c_f$ to $d_{\rm h}$,
$d_{\text f}$ and $\nu$,
\begin{equation}
  \displaystyle
  c_f= 1 + \frac{d_{\rm h}-1/\nu}{d_{\rm f}}\;,
  \label{eqtau}
\end{equation}
so that
\begin{equation}
\label{gouyet}
b=1-\frac{d_{\rm h}-1/\nu}{d_{\rm f}}\;,
\end{equation}
which leads $b=0.46$. This is also in accordance with the value we
observe, $b=0.47$.

\section{Conclusion}
\label{conc}

We have here studied the effect of the competition between disorder
and stress enhancement using the local load sharing fiber bundle
model.  We have done this by varying the disorder using a threshold
distribution controlled by a single-parameter $D$. When $D$ is
positive, the threshold distribution is a power law towards infinitely
weak elements, and when $D$ is negative, the distribution is a power
law towards infinitely strong elements. By defining an order parameter
distinguishing between the localized and non-localized phases, we find
two phase transitions, one at $D=D^+_c=0.47\pm 0.04$ and one at
$D=D^-_c=-1.03\pm 0.02$.  The transition for $D>0$ is second order and
the transition for $D<0$ is first order.  The second order transition
is governed by critical exponents that are consistent with
percolation, but also those found for the two-dimensional Ising model.
We then studied the spatiotemporal correlation function, finding the
same behavior as first seen by Furuberg et al.\ \cite{ffa88} in an
invasion percolation context. Some numerical values for the exponents
controlling the correlation function are different than found by
Furuberg et al., which, by following the scaling analysis of Roux and
Guyon \cite{rg89}, Maslov \cite{m95} and Gouyet \cite{g90}, we
successfully related to other exponents describing the geometry of the
cracks.

The conclusions we present here differ from those that Stormo et
al. \cite{sgh12} presented.  In that work, based on the {\it
  soft-clamp fiber bundle model,\/} the threshold distribution was the
uniform one on the unit interval, i.e.\ $D=1$.  The parameter that was
varied, was the elastic constant of the clamps to which the fibers are
connected.  For a given value of this parameter, the breakdown process
would proceed in the beginning as an uncorrelated percolation process
up to a certain point at which localization would set in.  From this
point on, the breakdown process would continue through the growth of a
single cluster of broken fibers.  This transition would {\it not\/} be
a phase transition, but a crossover.

It would be of great interest to repeat the analysis we have presented
here using the soft-clamp fiber bundle model.  Only then we will be
able to distinguish what is model dependent in our conclusions and
what is not.

\section*{Acknowledgment}
\label{Acknow}

The authors thank Eirik G. Flekk{\o}y for interesting
discussions. This work was partly supported by the Research Council of
Norway through its Centres of Excellence funding scheme, project
number 262644. SS was supported by the National Natural Science
Foundation of China under grant number 11750110430.



\end{document}